\documentclass{ws-procs9x6}
\newcommand{\open}{< \kern -0.3 em \scriptscriptstyle {)}}

\begin{document}

\title{Partonic structure of the Nucleon in QCD and Nuclear Physics: \\
new developments from old ideas}

\author{M.  RADICI}

\address{Istituto Nazionale di Fisica Nucleare, Sezione di Pavia, and \\
Dipartimento di Fisica Nucleare e Teorica, Universit\`a di Pavia, \\ 
via Bassi 6, \\
27100 Pavia, Italy  \\ 
E-mail: radici@pv.infn.it}

\maketitle

\abstracts{The nucleon is an ideal laboratory to solve QCD in the 
nonperturbative regime. There are several experimental observations that still lack a
rigorous interpretation; they involve the nucleon as a (polarized) target as 
well as a beam (in collisions and Drell-Yan processes). These data look like 
big azimuthal and spin asymmetries, related to the 
transverse polarization and momentum of the nucleon and/or the final detected 
particles. They suggest internal reaction mechanisms that are suppressed in collinear 
perturbative QCD but that are "natural" in Nuclear Physics: quark helicity flips,
residual final state interactions, etc.. In my talk, I will 
give a brief survey of the main results and I will flash the most recent 
developments and measurements.}

\section{Introduction}
\label{sec:intro}

The spin structure of the nucleon can be best studied by using the socalled 
spin asymmetry, defined as the ratio between the difference and the 
sum of differential cross sections obtained by flipping the spin of one of the
particles involved in the reaction. Spin asymmetries are known since a 
long time; historically, the first one was obtained at FERMILAB, where an 
anomalous large transverse polarization of the $\Lambda$ produced in 
proton-nucleon annihilations was measured~\cite{Bunce:1976yb}, surviving even at 
large values of the $\Lambda$ transverse momentum. More recently, similar 
anomalously large asymmetries have been observed, for example by the 
STAR~\cite{Adams:2003fx} collaboration in 
inclusive pion production from collisions of transversely polarized protons, as
well as in Deep-Inelastic Scattering (DIS) of lepton probes on polarized protons
by the HERMES~\cite{Airapetian:2004tw} and CLAS~\cite{Avakian:2003pk} collaborations. 

Assuming that partons are collinear with their parent hadrons and that a
factorization theorem exists for the process at hand, QCD relates the spin
asymmetry for transversely polarized objects to the off-diagonal components of
the elementary scattering amplitude in the parton helicity basis; actually, to
the imaginary part of products of such components~\cite{Kane:1978nd}. But with 
the above assumptions any mechanism flipping the helicity of the parton is suppressed 
in QCD up to terms proportional to its mass. Since we are considering here 
mainly protons, which contain only $up$ and $down$ valence quarks, we can assume 
that deviations to this rule are negligible. Therefore, QCD in collinear 
approximation is not capable to explain the above mentioned amount of 
experimental observations. 

Indeed, since in polarized DIS the structure function $g_{_T} = g_1 + g_2$
(related to the partonic transverse spin distribution) appears at subleading
twist, a common prejudice has always driven people to consider transverse spin
effects as suppressed because inextricably associated to off-shellness,
higher-order quark-gluon interactions, etc.. But in perturbative QCD (pQCD),
longitudinal- and transverse-spin effects can be described on the same footing
provided that the appropriate helicity and transverse basis are selected,
respectively~\cite{Jaffe:1996zw}. As a consequence, the spin structure of the
proton at leading twist is not fully exploited by just the well known momentum
and helicity distributions $f_1(x)$ and $g_1(x)$ (or, $q(x)$ and $\Delta q(x)$,
in another common notation). A third one, the transversity $h_1(x)$ (or, $\delta
q(x)$), is necessary which is basically unknown because it is related to
helicity-flip mechanisms, as it should be clear from the above discussion. Since
helicity and chirality coincide at leading twist, it is usually referred to as a
chiral-odd distribution. Observables, like the cross section, must be
chiral-even; hence, the transversity needs another chiral-odd partner to be
extracted from a spin asymmetry. This is why, for example, it is suppressed in
simple inclusive DIS (for a review on the transversity distribution, see
Ref.~\cite{Barone:2003fy}). 

From this short introduction, it should hopefully be clear that the standard
framework in which pQCD is usually calculated, namely collinear approximation
neglecting transverse spin components, is not adequate to interpret the wealth of
data collected over the years under the form of spin (azimuthal) asymmetries. In 
the following, I will review the main flowing ideas about improving such
framework.

\section{Intrinsic transverse momentum distribution of partons}
\label{sec:pT}

Let consider the annihilation process $p p^\uparrow \rightarrow \pi X$, where a pion is
semi-inclusively detected. If a factorization proof holds for the elementary process, 
in collinear approximation the differential cross section can be schematically 
written as
\begin{eqnarray}
\frac{d\sigma}{d\eta \, dP_\pi \, d\phi_{_S}} &\propto &\sum_{abcd} \int dx_a dx_b dz_c 
dz_d \, \phi(x_a,Q^2)\, \phi(x_b,Q^2) \, d\hat\sigma (ab^\uparrow \rightarrow 
c^\uparrow d) \nonumber \\
& &\qquad \times \chi (z_c,Q^2)\, \delta(z_c - \bar{z}_c)\, \delta(z_d - 1)  \; ,
\label{eq:collinear}
\end{eqnarray}
where the functions $\phi$ are the distributions for the two annihilating
partons carrying the fractions $x_a$ and $x_b$ of the two corresponding proton
momenta, respectively, while $\chi$ is the fragmentation function for the pion with 
momentum $P_\pi$ and carrying a fraction $z_c$ of the fragmenting parton. Momentum 
conservation at the partonic level implies that $z_c$ is constrained to $\bar{z}_c$, 
which is a function of $x_a, x_b,$ of the center-of-mass (cm) energy $s$ and of the 
rapidity $\eta$. The elementary cross section $d\hat\sigma$ 
with transversely polarized partons can be deduced from Ref.~\cite{Gastmans:1990xh}. 
The angle $\phi_{_S}$ is formed between the directions of the polarization and of 
the momentum of the polarized proton. 

If this framework is not appropriate to describe experimental observations of spin
asymmetries involving transversely polarized objects, a possible generalization
consists in assuming (and trying to prove) that the factorization holds also when an
explicit dependence upon the parton transverse momenta is introduced in
Eq.~(\ref{eq:collinear}), namely
\begin{eqnarray}
\frac{d\sigma}{d\eta \, dP_\pi \, d\phi_{_S}} &\propto &\sum_{abcd} \int dx_a dx_b dz_c 
dz_d d{\bf p}_{Ta} d{\bf p}_{Tb} d{\bf p}_{T\pi} \, \delta(z_c - \bar{z}_c)\, \delta 
({\bf p}_{T\pi} \cdot \hat{\bf p}_c) \nonumber \\
& &\qquad \times \phi (x_a,{\bf p}_{Ta},Q^2)\, \phi (x_b,{\bf p}_{Tb},Q^2) \, 
d\hat\sigma (a b^\uparrow \rightarrow c^\uparrow d) \nonumber \\
& &\qquad \times \chi (z_c,{\bf p}_{T\pi},Q^2)\, \delta(z_d - 1) \; .
\label{eq:conpT}
\end{eqnarray}
At present, this factorization scheme has been proven only for Drell-Yan and $e^+ e^-$
processes~\cite{Coll-Sop-Ster}, as well as for semi-inclusive DIS in some kinematical
regions~\cite{Ji-Ma-Yuan}. Therefore, for the considered process Eq.~(\ref{eq:conpT})
is no more than an assumption, but it leads to several interesting consequences.

\begin{figure}[ht]
%\epsfxsize=10cm   %width of figure - will enlarge/reduce the figures
%\epsfbox{fig3.eps}
%\figurebox{2cm}{3cm}{} %to have a box alone 
\begin{center}
\epsfxsize=2.5cm \epsfbox{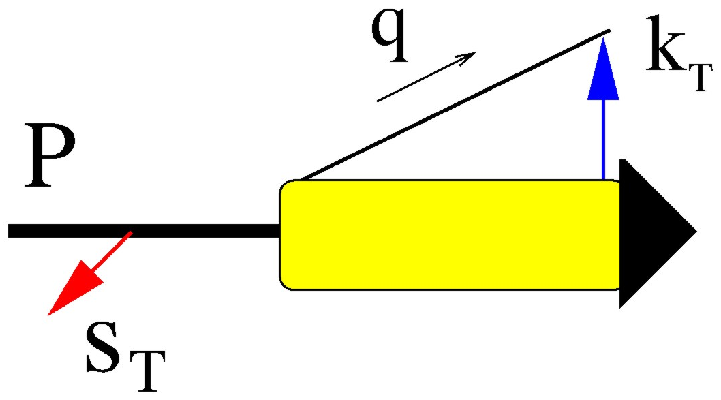} \hspace{.5cm} \epsfxsize=4cm 
\epsfbox{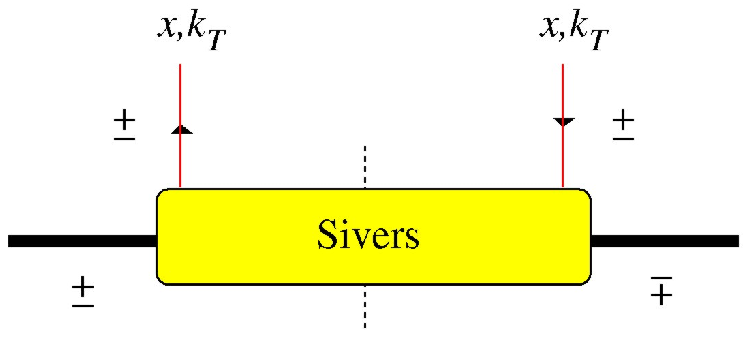}  \\[-0.5cm]
\epsfxsize=2.5cm \epsfbox{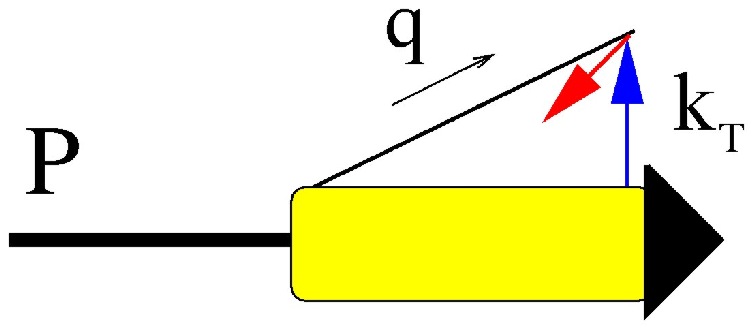} \hspace{.5cm} \epsfxsize=4cm 
\epsfbox{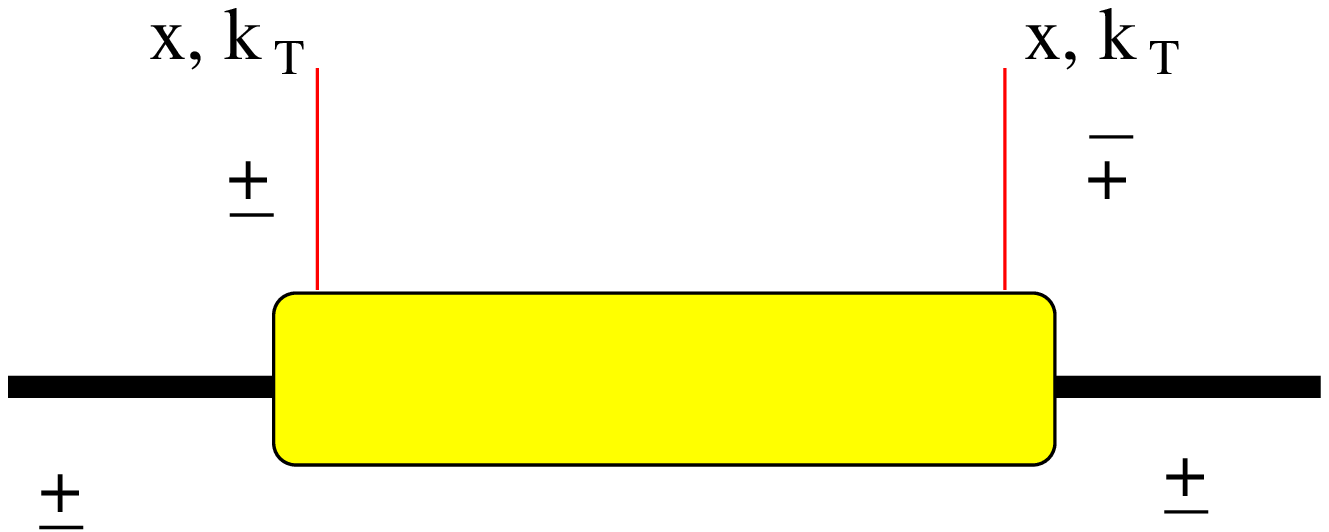} \\
\epsfxsize=2.5cm \epsfbox{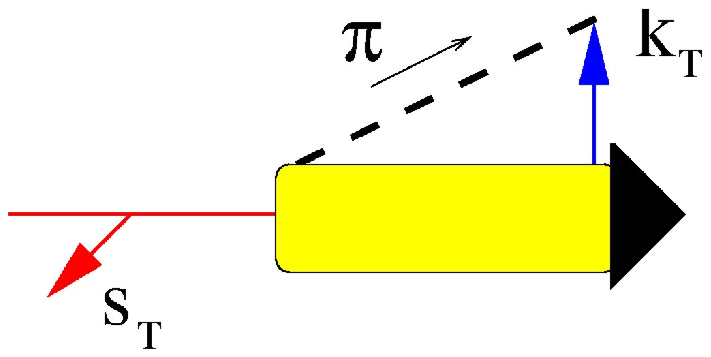} \hspace{.5cm} \epsfxsize=3cm 
\epsfbox{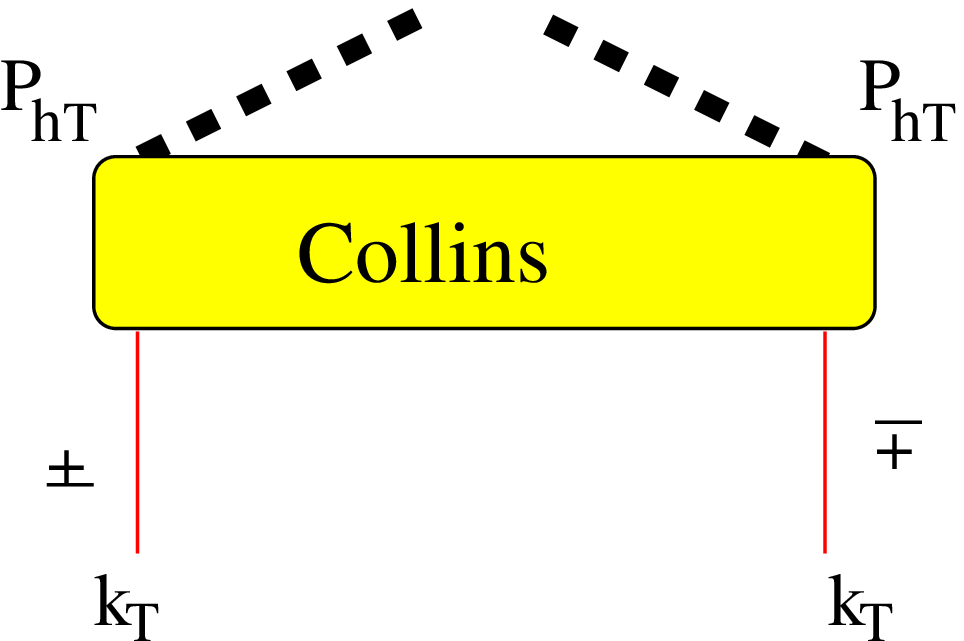} \\
\epsfxsize=2.5cm \epsfbox{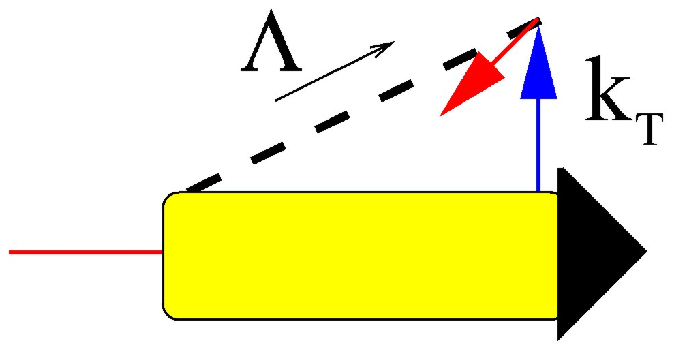} \hspace{.5cm} \epsfxsize=3cm 
\epsfbox{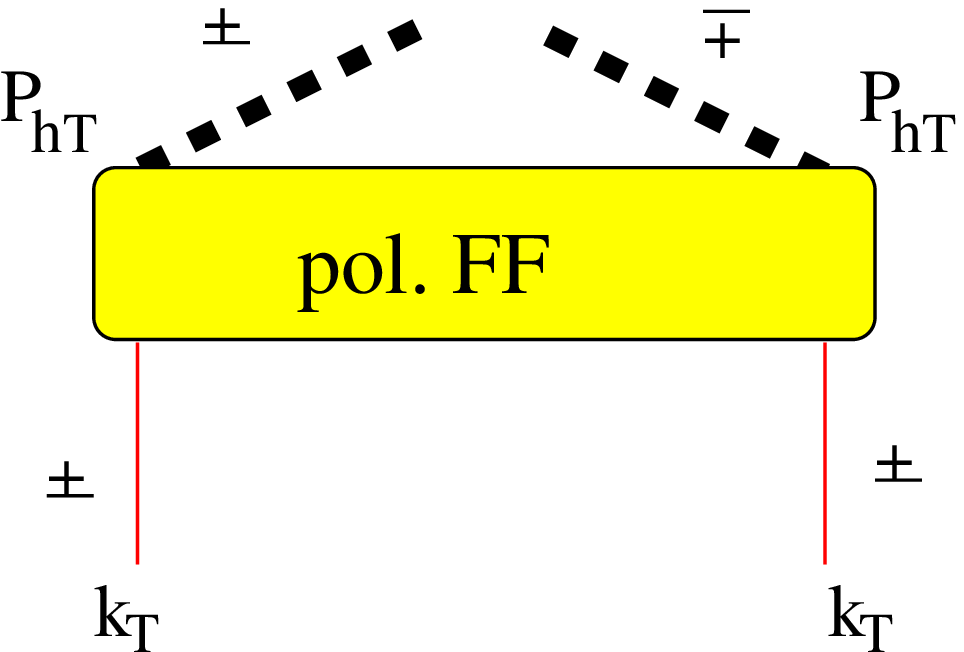}
\end{center}
\caption{On the left, from top to bottom: a- the nonperturbative correlation giving 
rise to the Sivers effect; b- the same for transversely polarized quarks in
unpolarized protons; c- the same for the Collins effect; d- the same for the
polarized fragmentation. On the right, from top to bottom, the corresponding 
field-theoretical interpretations.
\label{fig:f1Tperp}}
\end{figure}

When partons have an intrinsic transverse momentum with respect to the direction of the
parent hadron momentum, the list of the leading-twist distribution and fragmentation
functions is far longer than in the collinear case~\cite{Muld-Tang}. Several
(chiral-odd) functions appear with a specific number density interpretation, that can
originate new interesting spin effects. In the initial state, when the proton is 
transversely polarized, we have
\begin{equation}
f(q/p^\uparrow) = f_1^q(x,{\bf p}_{_T}) - f_{1T}^{\perp\,q}(x,{\bf p}_{_T})\, 
\frac{\hat{\bf P}\times {\bf p}_{_T} \cdot {\bf S}_{_T}}{M} \; ,
\label{eq:sivers}
\end{equation}
where the function $f_{_{1T}}^\perp (x,{\bf p}_{_T}) \propto f(q/p^\uparrow) - 
f(q/p^\downarrow)$ describes how the distribution gets
distorted by the proton transverse polarization. This is possible because
$f_{_{1T}}^\perp$ appears weighted by the correlation ${\bf P} \times {\bf p}_{_T} 
\cdot {\bf S}_{_T}$ between the momentum ${\bf P}$ and transverse spin ${\bf S}_{_T}$ 
of the proton with mass $M$, and the transverse momentum ${\bf p}_{_T}$ of the parton 
(see Fig.~\ref{fig:f1Tperp}-a). As a consequence, the final pion emerging from the
fragmentation can be deflected into different directions according to the 
transverse polarization of the initial proton, the socalled Sivers 
effect~\cite{Sivers:1990cc}. From the field-theoretical point of view,
$f_{_{1T}}^\perp$ can be represented as in Fig.~\ref{fig:f1Tperp}-a, i.e. it is
diagonal in the parton helicity basis but not in the hadron one. As such, it is
chiral-even but does not fullfil the constraints of time-reversal invariance (in
jargon, it is T-odd); a possible interpretation for the latter feature is that in the
considered process $p p^\uparrow \rightarrow \pi X$, a sort of initial state
interaction occurs before the collision~\cite{Anselmino:2004ky}, which prevents the 
$S$ matrix from being invariant under time-reversal transformations. Interestingly,
$f_{_{1T}}^\perp$ can be linked to the helicity-flip Generalized Parton Distribution
(GPD) $E$, where the correlation between ${\bf S}_{_T}$ and ${\bf p}_{_T}$ can be 
directly interpreted as due to the orbital angular momentum of the partons 
themselves~\cite{Burk}.

When the proton is not polarized, we can have
\begin{equation}
f(q^\uparrow/p) = \frac{1}{2}\,\left( f_1^q(x,{\bf p}_{_T}) - 
h_1^{\perp\,q}(x,{\bf p}_{_T}) \, 
\frac{\hat{\bf P}\times {\bf p}_{_T} \cdot {\bf S}_{_{qT}}}{M}
\right) \; ,
\label{eq:boer}
\end{equation}
where the function $h_1^\perp (x,{\bf p}_{_T}) \propto f(q^\uparrow /p) -
f(q^\downarrow/p)$ describes the influence of the parton transverse polarization 
${\bf S}_{_{qT}}$ on its momentum distribution inside an unpolarized proton.
Alternatively to the Sivers effect, the final pion can be deflected according to the 
direction of ${\bf S}_{_{qT}}$ entering the correlation factor (see 
Fig.~\ref{fig:f1Tperp}-b). In the corresponding handbag diagram, the function is 
diagonal in the hadron helicity but not in the parton one, hence it is chiral-odd and it 
represents a natural partner of transversity for the considere process~\cite{Boer}. 
Extraction of $h_1^\perp$ is of great importance, because it is believed to be 
responsible for the well known violation of the Lam-Tung sum 
rule~\cite{Conway}, an anomalous big azimuthal asymmetry of the 
unpolarized Drell-Yan $p p \rightarrow \mu^+ \mu^- X$ process, that neither 
Next-to-Leading Order (NLO) QCD calculations~\cite{Bran-Nach-Mirk}, nor higher twists or 
factorization-breaking terms in NLO QCD~\cite{Bran-Brod,Hoy-Vant,Berg-Brod} are able to 
justify.

In the final state, a transversely polarized/unpolarized parton can fragment into a
hadron with mass $M_h$ and carrying a fraction $z$ of the momentum. As for the former,
since we are considering a semi-inclusive process with a pion in the final state, we can
have
\begin{equation}
D(h/q^\uparrow) = D_1^q(z,{\bf P}_{_{hT}}) + H_1^{\perp\,q}(z,{\bf P}_{_{hT}})\, 
\frac{\hat{\bf p}\times {\bf P}_{_{hT}} \cdot {\bf S}_{_{qT}}}{zM_h} \; ,
\label{eq:collins}
\end{equation}
where $D_1^q(z,{\bf P}_{_{hT}})$ is the probability for a quark $q$ to
fragment into a hadron with transverse momentum ${\bf P}_{_{hT}}$, while 
$ H_1^{\perp\,q}(z,{\bf P}_{_{hT}}) \propto D(h/q^\uparrow ) - D(h/q^\downarrow)$ is the
analogue for a quark with transverse polarization ${\bf S}_{_{qT}}$, i.e. the Collins
function~\cite{collins}. Again, also the polarization of the fragmenting quark can be 
responsible for an asymmetric distribution of the detected pion via the correlation 
$\hat{\bf p}\times {\bf P}_{_{hT}} \cdot {\bf S}_{_{qT}}$ (see Fig.~\ref{fig:f1Tperp}-c). In
the same figure, the diagram corresponding to $H_1^\perp$ displays a helicity flip for
the fragmenting quark; the Collins function is both chiral-odd and T-odd, and it
represents another possible partner for extracting the transversity through the Collins
effect. Several experimental collaborations are pursuing this 
goal~\cite{Adams:2003fx,Airapetian:2004tw,Avakian:2003pk}, with the help of
theoretical calculations as well~\cite{efre-goe-schw,bacch-kund-}. 

Finally, the last combination is given by 
\begin{equation}
D(h^\uparrow/q) = \frac{1}{2}\,\left( D_1^q(z,{\bf P}_{_{hT}}) + 
D_{_{1T}}^{\perp\,q}(z,{\bf P}_{hT})\, \frac{\hat{\bf p}\times {\bf P}_{_{hT}} \cdot 
{\bf S}_h}{zM_h} \right) \; .
\label{eq:polff}
\end{equation}
The function $D_{_{1T}}^{\perp\,q}(z,{\bf P}_{hT}) \propto D(h^\uparrow/q) -
D(h^\downarrow/q)$ describes the fragmentation of an unpolarized parton into a
hadron with polarization ${\bf S}_h$ like, e.g., the $\Lambda$: it is not pertinent to 
the process we have here selected, but it is anyway important because the correlation 
$\hat{\bf p}\times {\bf P}_{_{hT}} \cdot {\bf S}_h$ is believed to be the mechanism
responsible for the observed asymmetric production of $\Lambda$ in unpolarized proton
collisions~\cite{anselm-boer,Anselmino:2004ky} (see Fig.~\ref{fig:f1Tperp}-d). From the related
diagram, the function $D_{_{1T}}^{\perp}$ is related to a helicity flip of just the
polarized hadron; it is then chiral even, but T-odd. 

\section{Interference fragmentation functions}
\label{sec:iff}

From previous section, it emerges that in the non-collinear factorized framework of
Eq.~(\ref{eq:conpT}) at least three different interpretations arise at leading twist for
the spin asymmetry observed in the $p p^\uparrow \rightarrow \pi X$ process: the Sivers
effect, related to the convolution $f_1^a \otimes f_{_{1T}}^{\perp b} \otimes D_1^c$ for
the elementary process $a + b \rightarrow c + d$; the Collins effect, related to the
convolution $f_1^a \otimes h_1^b \otimes H_1^{\perp c}$ for $a + b^\uparrow \rightarrow 
c^\uparrow +d$; the effect related to the convolution $h_1^{\perp a} \otimes h_1^b 
\otimes D_1^c$ for $a^\uparrow + b^\uparrow \rightarrow c+d$. An intense experimental 
and theoretical activity is ongoing in this field in order to unravel the physics
contained in the asymmetry data, and also in related topics, like for
example the new project of extracting transversity from Drell-Yan using (polarized)
antiprotons~\cite{pax}.

\begin{figure}[ht]
%\epsfxsize=10cm   %width of figure - will enlarge/reduce the figures
%\epsfbox{fig3.eps}
%\figurebox{2cm}{3cm}{} %to have a box alone 
\begin{center}
\epsfxsize=5cm \epsfbox{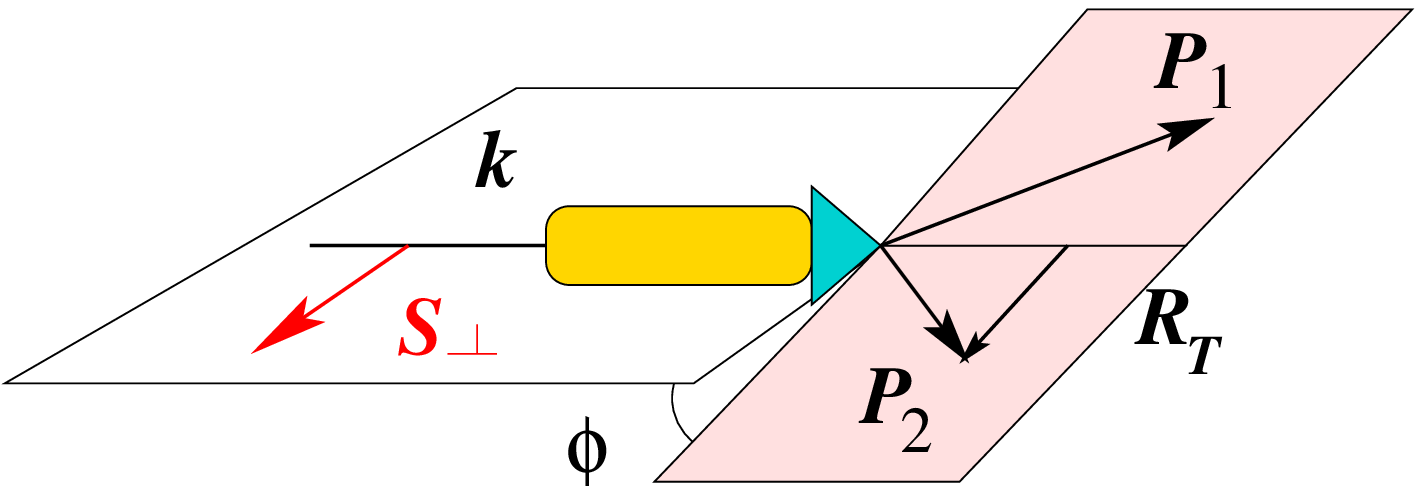}
\end{center}
\caption{The mechanism leading to a spin asymmetry for a transversely polarized quark
fragmenting into two unpolarized hadrons.
\label{fig:iff}}
\end{figure}

But all the above mechanisms require an explicit dependence of the distribution and
fragmentation functions upon an intrinsic tranverse momentum of partons. We already
stressed that in such a non-collinear framework the factorization leading to
Eq.~(\ref{eq:conpT}) has not yet been proven for hadron-hadron collisions. A lot of
work is being done also in this specific, though quite general,
subject~\cite{metz,boer-pijl-muld,coll-metz}. Nevertheless, it would be desirable to
find out a mechanism that leads to a spin asymmetry without invoking an explicit 
dependence on ${\bf p}_{_T}$. Since we are
considering here final states made of unpolarized hadrons only, we need an additional
4-vector that can be provided by semi-inclusively detecting a second hadron inside the
same jet of the first one. In fact, in this case the system of two unpolarized hadrons
with momenta $P_1$ and $P_2$ is described by two 4-vectors, the cm one, $P_h 
= P_1 + P_2$ and the relative one, $R = (P_1 -P_2)/2$, by which we can construct the
correlation ${\bf S}_{_{qT}} \cdot {\bf R} \times {\bf P}_h$~\cite{coll-lad,jaffe,noi1}
 (see Fig.~\ref{fig:iff}). 

Indeed, for the DIS lepton-induced production from a transversely polarized proton of 
two pions with the cm momentum aligned to the jet axis (i.e., with 
${\bf P}_{_{hT}} = 0$), it has been shown~\cite{noi2} that at leading twist the cross
section contains a term like $({\bf S}_{_{qT}} \times {\bf R}_{_T}) \, h_1(x) \, 
H_1^{\open}(z,M_h^2,{\bf R}_{_T}^2)$, where the chiral-odd (T-odd) $H_1^{\open}$ is
representative of a new class of fragmentation functions, the Interference Fragmentation
Functions (IFF). A spin asymmetry can be built by flipping the spin of the transversely
polarized proton target, which isolates the transversity $h_1$ at leading twist via
$H_1^{\open}$. A thorough analysis of IFF has been carried out up to
twist-3~\cite{noi-twist3}, and $H_1^{\open}$ is the only one surviving at leading twist
in the collinear situation, since it is related to the azimuthal position of the plane 
containing the two pions: the latter ones are differently distributed in the azimuthal
angle $\phi$ of Fig.~\ref{fig:iff} according to the direction of the transverse
polarization of the fragmenting quark. A partial-wave analysis of the two-pion system
allows to isolate the main channels, where the two pions are produced in a relative $s$
or $p$ wave. Interference between the two or between different components of $p$ waves
is responsible of the T-odd nature of IFF~\cite{noi-LM}. These functions can be 
extracted in principle from the process $e^+ e^- \rightarrow (\pi \pi)_{jet1} 
(\pi \pi)_{jet2} X$~\cite{noi-e+e-}. Measurements aiming at the extraction of IFF are 
under way at HERMES (DIS) and BELLE ($e^+ e^-$) and could be possible at COMPASS and 
BABAR too. However, it has been recently proposed~\cite{noi-pp} that a consistent 
extraction of $h_1$ and $H_1^{\open}$ could be achieved by considering again the 
collision of (un)polarized protons leading to one or two pion pairs semi-inclusively 
detected in the final state. For the case $p p^\uparrow \rightarrow (\pi \pi) X$, the 
cross section contains the convolution $f_1^a \otimes h_1^b \otimes H_1^{\open c}$ 
for the elementary process $a + b^\uparrow \rightarrow c^\uparrow d$. On the other 
hand, for the corresponding unpolarized process $p p \rightarrow (\pi \pi)_{jet1} 
(\pi \pi)_{jet2} X$ the cross section contains $f_1^a \otimes f_1^b \otimes 
H_1^{\open c} \otimes H_1^{\open d}$ for $a + b \rightarrow c^\uparrow + d^\uparrow$. 
By combining the two measurements, information on the two unknowns $h_1$ and
$H_1^{\open}$ can be extracted consistently in the same experiment. Finally, the very
same formalism offers the possibility of observing for the first time the transverse
linear polarization of gluons, even using spin-${\textstyle \frac{1}{2}}$
targets~\cite{noi-pp}.

%\section*{Acknowledgments}

\end{document}